\documentclass{camera}

\def \apj {ApJ}

\def \aap {A\&A}

\def \mnras {MNRAS}

\begin{document}

%
\title{The X-ray flaring emission from\\ High Mass X-ray Binaries:\\ the effects of wind inhomogeneities}

%
\author{L. Ducci$^{1,2}$, L. Sidoli$^{2}$, P. Romano$^{3}$, A. Paizis$^{2}$ \and S. Mereghetti$^{2}$}

%
\organization{$^{1}$ Dipartimento di Fisica e Matematica, Universit\`a degli Studi dell'Insubria, Via Valleggio 11, I-22100  Como, Italy\\
$^{2}$ INAF, Istituto di Astrofisica Spaziale e Fisica Cosmica, Via E. Bassini 15, I-20133 Milano, Italy\\
$^{3}$ INAF, Istituto di Astrofisica Spaziale e Fisica Cosmica, Via U. La Malfa 153, I-90146 Palermo, Italy}

\maketitle

\begin{abstract}
We have developed a clumpy stellar wind model for OB supergiants in order to compare
predictions of this model with the X-ray behaviour of
both classes of persistent and transient High Mass X-ray Binaries (HMXBs).
\end{abstract}

%

\section{Introduction}

The Galactic plane survey performed by \emph{INTEGRAL}
satellite in the last 7 years allowed the discovery of 
a new class of HMXBs with OB supergiants,
called \emph{Supergiant Fast X-ray Transients} (SFXTs):
they are transient sources which sporadically
exhibit flares, with duration of a few hours,
reaching luminosity of $10^{36}-10^{37}$~erg~s$^{-1}$,
and a high dynamic range, spanning 3 to 5 orders 
of magnitude \cite{Sguera2005}.
The accretion mechanism responsible for the peculiar SFXT
behaviour is still not clear (see \cite{Sidoli2009} and 
references therein for a recent review).
Among the different possible explanations, \cite{intZand2005} proposed 
that the SFXTs flares are produced
by the accretion of dense blobs of matter from the companion wind.

\section{A new clumpy stellar wind model}

Since the accretion of one single clump of matter
cannot explain the observed X-ray lightcurve in SFXTs
in outburst (see e.g. \cite{Romano07, Sidoli07}),
we developed a more detailed clumpy 
stellar wind model for OB supergiants in HMXBs \cite{Ducci2009}.
For the first time, a mass and radius distributions for the clumps
have been introduced.
We assume for the clump the same velocity law of a smooth stellar
wind, as suggested by \cite{Lepine2008},
and following  \cite{Howk2000}, 
we find that the clump size increases with the distance 
from the supergiant star \cite{Ducci2009}.
For each mass of the clump, we derived the upper and 
lower-limits for the clump radius.
We also modified the Bondi-Hoyle accretion model to take into account
the presence of inhomogeneities in the wind.
In this way, we were able to compare the observerd lightcurves of
SFXTs and of persistent HMXBs
with the calculated ones.
We found that the observational characteristics of the flares,
luminosity, duration, number of flares produced, 
do not depend only on the orbital parameters, 
but are also significantly affected 
by the properties of the clumps \cite{Ducci2009}.
This model has been successfully applied to four HMXBs:
Vela~X$-$1, 4U~1700$-$37, IGR~J18483$-$0311 and IGR~J11215$-$5952 \cite{Ducci2009, Romano2009}.

\paragraph{Acknowledgments}
L.D. thanks Prof. A. Treves for very helpful discussions.
This work was supported by ASI contracts  I/023/05/0,
I/088/06/0 and I/008/07/0.

%
\end{document}